%% file: main.tex
\documentclass{article}

\usepackage{comment}

\usepackage{microtype}
\usepackage{booktabs} 

\usepackage{hyperref}

\usepackage{xcolor}
\usepackage{amsmath}
\usepackage{filecontents}
\usepackage{enumitem}
\usepackage{pbox}
\usepackage{graphicx}
\usepackage{subfig}

\usepackage{tablefootnote}
\usepackage[justification=centering]{caption}

\graphicspath{{images/}}

\usepackage{appendix}
\usepackage{amsthm}
\usepackage{float}
\usepackage{amsfonts}

\usepackage{algorithmic}

\newcommand{\name}{DiffusionPipe}

\usepackage{makecell}
\usepackage{hhline}
\usepackage{array,multirow}

\RequirePackage[normalem]{ulem} 
\RequirePackage{color}\definecolor{RED}{rgb}{1,0,0}\definecolor{BLUE}{rgb}{0,0,1} 

\usepackage[accepted]{mlsys2024}

\mlsystitlerunning{DiffusionPipe: Training Large Diffusion Models with Efficient Pipelines}

\begin{document}

\twocolumn[
\mlsystitle{DiffusionPipe: Training Large Diffusion Models with Efficient Pipelines} 



\mlsyssetsymbol{equal}{*}
\mlsyssetsymbol{aws_intern}{*}


\begin{mlsysauthorlist}
\mlsysauthor{Ye Tian}{hku,aws_intern}
\mlsysauthor{Zhen Jia}{aws}
\mlsysauthor{Ziyue Luo}{osu}
\mlsysauthor{Yida Wang}{aws}
\mlsysauthor{Chuan Wu}{hku}
\end{mlsysauthorlist}

\mlsysaffiliation{hku}{The University of Hong Kong, Hong Kong}
\mlsysaffiliation{aws}{Amazon Web Services, USA}
\mlsysaffiliation{osu}{The Ohio State University, USA}

\mlsyscorrespondingauthor{Ye Tian}{yetiansh@connect.hku.hk}

\mlsyskeywords{Machine Learning, MLSys}

\vskip 0.3in


\input{sources/abstract}

]



\printAffiliationsAndNotice{\textsuperscript{*}Work done during internship at AWS.}

\input{sources/introduction}

\input{sources/background}

\input{sources/system_overview}

\input{sources/trainable}

\input{sources/pipeline_bubble_filling}

\input{sources/evaluation}

\input{sources/conclusion}

\input{sources/acknowledgement}

\bibliography{reference}
\bibliographystyle{mlsys2024}




\end{document}

%% file: sources/abstract.tex
\begin{abstract}
Diffusion models have emerged as dominant performers for image generation. 
To support training large diffusion models, this paper studies pipeline parallel training of diffusion models and proposes~\name{}, a synchronous pipeline training system that advocates innovative pipeline bubble filling technique, catering to structural characteristics of diffusion models.
State-of-the-art diffusion models typically include trainable (the backbone) and non-trainable (e.g., frozen input encoders) parts. 
We first unify optimal stage partitioning and pipeline scheduling of single and multiple backbones in representative diffusion models with a dynamic programming approach.
We then propose to fill the computation of non-trainable model parts into idle periods of the pipeline training of the backbones by an efficient greedy algorithm, thus achieving high training throughput.
Extensive experiments show that~\name{} can achieve up to 1.41x speedup over pipeline parallel methods and 1.28x speedup over data parallel training on popular diffusion models.
\end{abstract}

%% file: sources/introduction.tex
\section{Introduction}\label{sec:introduction}
Diffusion models have become the dominant choice for content generation today, including text-image synthesis~\cite{choi2021ilvr} and video generation~\cite{ramesh2022hierarchical}.
Large diffusion models such as Stable Diffusion~\cite{rombach2022high}, ControlNet~\cite{zhang2023adding}, and Imagen~\cite{saharia2022photorealistic} achieve state-of-the-art performance in various scenarios.
There is a continuing trend to develop larger diffusion models by increasing the backbone size~\cite{rombach2022high,peebles2022scalable,bao2023all,podell2023sdxl}, cascading multiple backbones to enable higher resolution image generation~\cite{nichol2021glide, peebles2022scalable,saharia2022photorealistic,ho2022cascaded,podell2023sdxl}, and combining different transformer architectures with diffusion models~\cite{peebles2022scalable,zhang2023adding,wu2023visual}.

Data parallelism is adopted for distributed diffusion model training~\cite{falcon2019lightning,bian2021colossal,platen2022diffusers}.
For large diffusion models, this method duplicates parameters, 
which limits the training batch size~\cite{rombach2022high,ho2022cascaded,saharia2022photorealistic} and device utilization, and causes significant synchronization overhead, especially when the training scale is large~\cite{narayanan2019pipedream}.

Pipeline parallelism \cite{huang2019gpipe,narayanan2019pipedream,luo2022efficient} has been widely adopted to train large DNN models, 
which partitions networks across multiple devices and pipelines micro-batch processing across model partitions, substantially alleviating memory consumption on a single device and enabling larger training batch sizes. 
Although pipeline parallelism is potentially useful in enabling larger diffusion model training, it has not been well explored for diffusion models and its application faces several challenges as follows:

{\em First}, the structural characteristics and special training procedures of diffusion models cannot be handled well by traditional pipelining methods.
A diffusion model typically contains a trainable part with one or multiple backbone models (e.g., U-Net)~\cite{rombach2022high}, and a non-trainable part with frozen text and image encoders, and they are usually trained with special techniques such as self-conditioning~\cite{chen2022analog}, which involves an additional forward computation pass on the backbone.
Pipeline training involves only the trainable part, while the non-trainable part is not readily handled by existing pipeline training methods because it does not require pipelining.
Self-conditioning is beyond the scope of existing pipeline systems, as they assume that there is only one forward pass.

{\em Second}, pipeline bubbles are often significant in synchronous pipeline training~\cite{huang2019gpipe,fan2021dapple,luo2022efficient}, which is more widely used in practice due to not altering model performance but involves periodic pipeline flushing.
We identify a unique opportunity to fill the pipeline bubbles using the computation of non-trainable model components, to substantially improve device utilization and expedite training speed.
However, there are dependencies between the trainable and non-trainable part that block pipeline bubble filling by overlapping their execution.
In addition, how to partition the non-trainable part into sets of layers and insert them into the pipeline bubble is not studied.

{\em Third}, Non-trainable layers with extra-long execution time are common in frozen encoders~\cite{kingma2013auto}.
Such layers may not fit into any pipeline bubble and block filling pipeline bubble with all subsequent layers in the non-trainable part, which cannot be solved by only partitioning the non-trainable part into sets of layers.
In addition, as non-trainable layers' execution time is discrete, it is unlikely to fully utilize idle time in individual pipeline bubble, leading to performance degradation.

In this paper, we propose~{\em \name{}}, an efficient pipeline training system designed specifically for large diffusion models.
~\name{} systematically determines optimized model partitioning, stages, and replication settings while applying pipeline bubble filling techniques. 
These optimizations are tailored for a variety of representative diffusion models and training methods.
To the best of our knowledge, we are the first to enable efficient pipeline parallel training of diffusion models.
Our contributions can be summarized as follows:

$\triangleright$ We propose a unified dynamic programming-based algorithm for optimized model partitioning, that can handle various training scenarios, e.g., models with different numbers of backbones and models trained with self-conditioning.
The proposed partitioning algorithm optimizes the model partitioning scheme under various settings of number of stages and number of micro-batches, with performance comparable to state-of-the-art pipeline paradigms under traditional pipelining, and effectively handles scenarios beyond traditional pipelining and specific to diffusion models.

$\triangleright$ We design a novel pipeline bubble filling strategy that fills the non-trainable part computation into the bubble time of the pipeline training of the backbone(s), effectively eliminating pipeline bubbles.
It efficiently partitions the non-trainable components and the input data for bubble filling, and effectively addresses dependencies between the non-trainable part and the trainable part by allowing \textit{cross-iteration} overlapping of backbone training of an iteration and non-trainable part computation of the next iteration and filling pipeline bubbles of the former with the latter.

$\triangleright$ We effectively handles extra-long non-trainable layers which do not fit into individual pipeline bubbles, by a \textit{partial-batch} processing design, for the non-training layer to process only a portion of a training batch.
Partial-batch layer's execution time can be precisely controlled by its input batch size, enabling it to be inserted into bubbles.
In addition, partial-batch layers help eliminate the remaining idle time in pipeline bubbles after inserting non-trainable layers (processing a complete batch).

We implement~\name{} and compare it to state-of-the-art data parallel training systems~\cite{rasley2020deepspeed} and ZeRO-3~\cite{rajbhandari2021zero}, together with synchronous pipeline training paradigms, including SPP~\cite{luo2022efficient} and GPipe~\cite{huang2019gpipe}.
Experimental results show that~\name{} achieves up to 1.28x speedup over data parallel training and up to 1.41x speedup over existing pipeline parallel methods on representative diffusion models.
We observe that~\name{} achieves almost complete elimination of pipeline bubbles and effectively handles multiple training scenarios of diffusion models.

%% file: sources/background.tex
\section{Background and Motivation}
\subsection{Diffusion models and training}

\begin{figure}[t]
    \centering
    \includegraphics[width=0.91\linewidth]{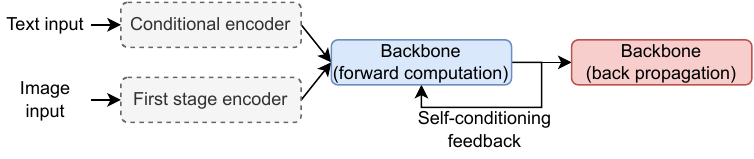}
    \vspace{-1em}
    \caption{Training process of Stable Diffusion v2.1~\cite{rombach2022high} and additional feedback of self-conditioning~\cite{chen2022analog}. Non-trainable components are marked in grey boxes.}
    \label{fig:training_process}
    \vspace{-15pt}
\end{figure}

\begin{table}[t]
\begin{center}
\begin{small}
\caption{Ratio of the forward time of the non-trainable part to the forward and backward time of the trainable part on A100 GPU}
\begin{tabular}{|c|c|c|c|c|}
\hline
\textbf{Model / Batch size} & \textbf{8} & \textbf{16} & \textbf{32} & \textbf{64} \\
\hhline{|=|=|=|=|=|}
Stable Diffusion v2.1 & 38\% & 41\% & 43\% & 44\% \\ \hline
ControlNet v1.0 & 76\% & 81\% & 86\% & 89\% \\ \hline
\end{tabular}
\label{tab:ratio_non_trainable_to_trainable}
\vspace{-20pt}
\end{small}
\end{center}
\end{table}

Diffusion models~\cite{ho2020denoising,song2020denoising,chen2022analog,rombach2022high,ho2022cascaded,saharia2022photorealistic,podell2023sdxl} are generative models that learn to reverse the diffusion process that gradually turns data into noise.
They typically comprise a backbone model that performs image generation and multiple frozen encoders that encode image and conditional information, e.g., class information~\cite{yu2015lsun}, text description~\cite{deng2009imagenet}, canny edge~\cite{canny1986computational} and human pose~\cite{kreiss2021openpifpaf}, and provide it as input to the backbone.
During diffusion model training, the encoders are typically fixed and executed in advance in the forward computation pass (referred to as the non-trainable part), while the backbone (the trainable part) is trained with both forward computation and backward propagation (Fig.~\ref{fig:training_process}).
Table~\ref{tab:ratio_non_trainable_to_trainable} compares the execution time of the non-trainable part and the training time (forward and backward) of the trainable part.

Some diffusion models, e.g., Cascaded Diffusion Models (CDM)~\cite{ho2022cascaded,ramesh2022hierarchical,podell2023sdxl}, involve multiple backbones of different capacities for high-resolution image generation. 
Multiple backbones accept the same encoder outputs, and each backbone also takes the output of the preceding backbone as input.
The training of backbones in a CDM are typically independent, and each is trained on a different set of devices using the same procedure, as shown in Fig.~\ref{fig:training_process}.

In the current mainstream diffusion models, U-Net~\cite{ho2020denoising,rombach2022high} is widely used as the backbone model. Transformer models can also serve as the backbone~\cite{peebles2022scalable,bao2023all}. 
T5-xxl~\cite{raffel2020exploring}, BERT~\cite{devlin2018bert} and CLIP~\cite{radford2021learning} text encoders are popular text encoders, while the image encoders are often variational auto-encoders~\cite{kingma2013auto}, ViT~\cite{dosovitskiy2020image} and CLIP image encoders.
There are corresponding encoders~\cite{zhang2023adding} for other modalities, such as canny edge and human pose.

Self-conditioning~\cite{chen2022analog} has become a very popular technique for training diffusion models~\cite{rombach2022high,saharia2022photorealistic,yuan2022seqdiffuseq}, which improves the sampling quality by introducing an additional forward computation pass of the backbone (Fig.~\ref{fig:training_process}).
The output of this forward pass is fed back to the backbone and serves as a conditional input.
The fidelity of the image is then improved because each step is conditioned on the previously generated samples.

\subsection{Pipeline parallel training, schedule and pipeline bubble}\label{sec:background_pipelining}
\begin{figure}[t]
    \centering
    \includegraphics[width=0.85\linewidth]{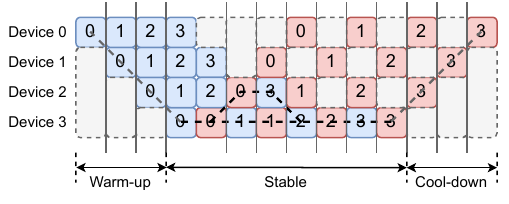}
    \vspace{-1em}
    \caption{FIFO-1F1B schedule of a DNN. Gray blocks without numbers indicate pipeline bubbles. Potential critical paths are marked with a dashed line. Numbers indicate micro-batch index in both forward (blue) and backward (pink) steps.}
    \label{fig:fifo_ofob_schedule}
    \vspace{-25pt}
\end{figure}

\begin{figure}[t]
    \centering
    \includegraphics[width=0.99\linewidth]{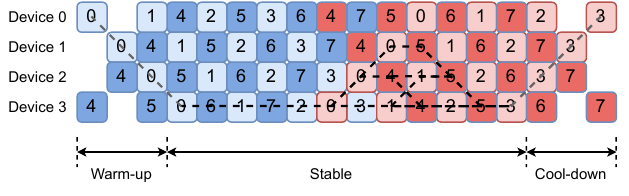}
    \vspace{-1em}
    \caption{Bidirectional pipeline schedule of a DNN. Communication omitted. The same meaning of number and color with Fig.~\ref{fig:fifo_ofob_schedule}. Micro-batch 0 to 3 pipeline from device 0 to 3 (down direction), while micro-batch 4 to 7 pipeline from device 3 to device 0 (up direction).}
    \label{fig:bidirectional_schedule}
    \vspace{-25pt}
\end{figure}

Pipeline parallel training partitions the model into stages, and each stage is deployed on a single device; the input data batch in each training iteration is divided into multiple micro-batches, which are processed through the model stages in a pipelined manner.
The micro-batch execution pipelines are typically scheduled by a First-In-First-Out (FIFO) heuristic~\cite{chen2015mxnet,abadi2016tensorflow,sergeev2018horovod}, which executes micro-batches on model stages according to their ready order.
The One-Forward-One-Backward (1F1B) schedule is widely adopted with FIFO, that alternatively executes forward computation and back propagation of micro-batches on each model stage in the stable phase of the pipeline execution (when multiple micro-batches are available to run on a model stage at the same time). As illustrated in Fig.~\ref{fig:fifo_ofob_schedule}, this schedule allows releasing intermediate activations and reduces peak memory usage by launching the backward computation when forward computation of the micro-batch is complete.

Chimera~\cite{li2021chimera} proposes bidirectional pipelining to reduce pipeline bubbles while retaining training synchronous.
Chimera maintains two pipelines of micro-batch execution in different device rank orders (i.e., pipeline directions) on the same set of model stages, with the two pipeline execution schedules being symmetric along the device dimension.
An example of bidirectional pipelining is shown in Fig.~\ref{fig:bidirectional_schedule}.
Each micro-batch's execution can fit perfectly into pipeline bubbles of its counterpart in the pipeline of the other direction (when the number of stages is even).

In synchronous pipeline training, pipeline bubbles generally exist in the pipeline schedule (Fig.~\ref{fig:fifo_ofob_schedule}).
There is a barrier that gradient synchronization imposes between pipeline stages of the trainable part of diffusion models at different iterations, disabling pipeline bubbles be filled by the trainable part at different iterations.
Therefore, although pipeline bubbles can be partially reduced by applying a better model partitioning and pipeline schedule, e.g., SPP~\cite{luo2022efficient} and Chimera~\cite{li2021chimera}, such approaches cannot fundamentally eliminate pipeline bubbles, as they only manipulate the trainable part of the model and do not take the non-trainable part into consideration.

\begin{table}[t]
\begin{center}
\begin{small}
\vspace{-1em}
\caption{Proportion of synchronization in training iteration time at local batch size 8 on A100 GPUs}
\begin{tabular}{|c|c|c|c|c|}
\hline
\textbf{Model / GPU count} & \textbf{8} & \textbf{16} & \textbf{32} & \textbf{64} \\
\hhline{|=|=|=|=|=|}
Stable Diffusion v2.1 & 5.2\% & 19.3\% & 36.1\% & 38.1\% \\ \hline
ControlNet v1.0 & 6.9\% & 22.7\% & 39.1\% & 40.1\% \\ \hline
\end{tabular}
\label{tab:synchronization_overhead}
\vspace{-25pt}
\end{small}
\end{center}
\end{table}

\subsection{Synchronization overhead and memory consumption of data parallel training}
\label{sec:observation_data_parallelism}
Diffusion models are largely trained using data parallelism nowadays~\cite{rombach2022high,ho2022cascaded,saharia2022photorealistic,podell2023sdxl}, which involves significant parameter synchronization overhead among devices and large memory consumption on each device that restricts the maximum feasible local batch size and the device utilization.
For example, Stable Diffusion is trained at a \textit{local} batch size of only 8 on each TPU-v3 (32GB) in~\cite{rombach2022high} consuming about 24.3 GB memory, which results in limited device utilization and exacerbates the synchronization portion of the training time.
The synchronization overhead in Table~\ref{tab:synchronization_overhead} is computed as the ratio of parameter synchronization time to the end-to-end time of a training iteration.
As the number of devices increases, parameter synchronization soon takes up a significant portion of the iteration time.
In summary, the data parallel style of diffusion model training limits the training batch size and imposes high synchronization overhead.

\subsection{Efficient pipeline bubble filling with non-trainable components}
\begin{figure}[t]
    \centering
    \subfloat[Stable Diffusion v2.1]{\includegraphics[width=0.45\linewidth]{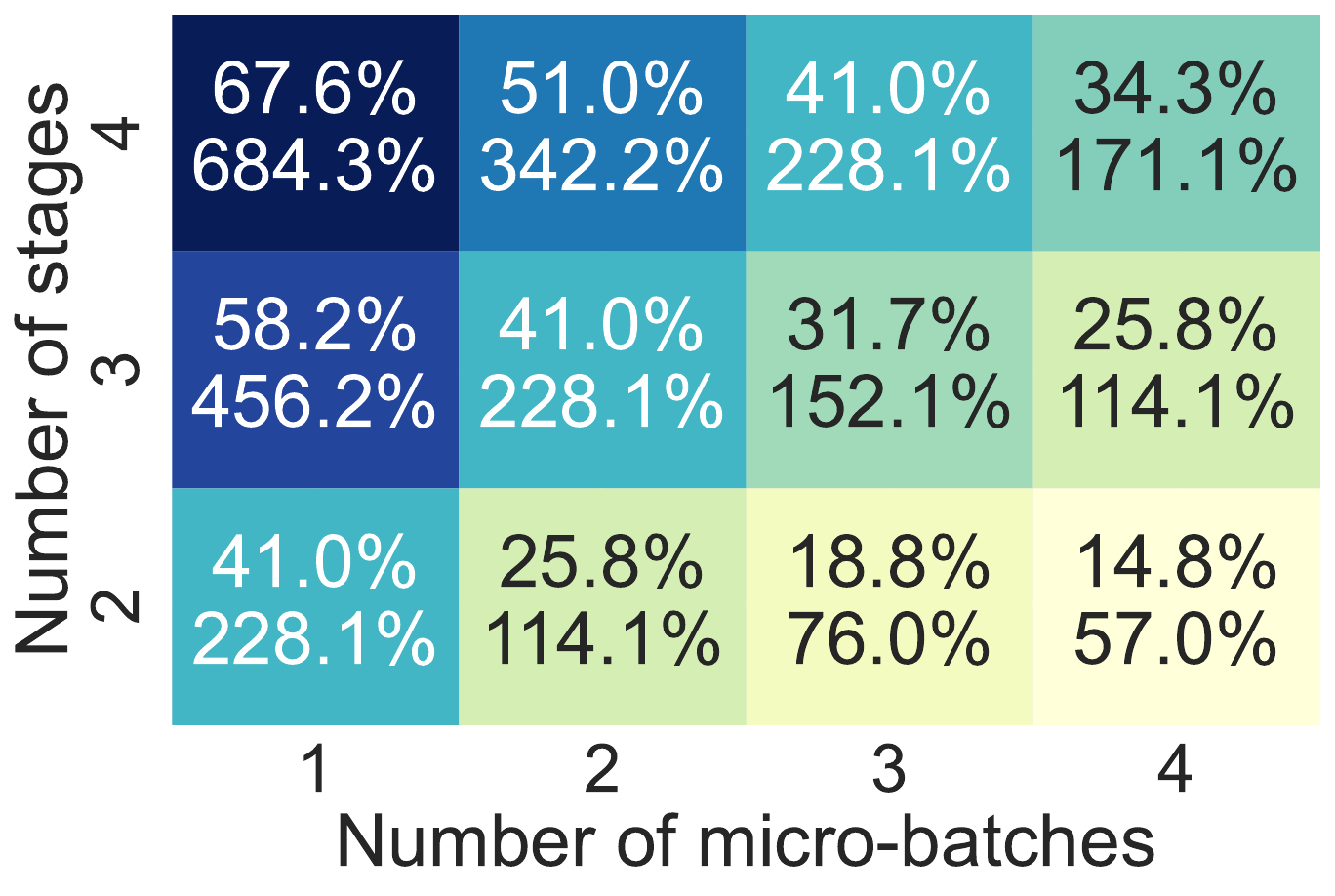}\label{fig:sd_pipeline_bubble_ratio}}\hskip1ex
    \subfloat[ControlNet v1.0]{\includegraphics[width=0.45\linewidth]{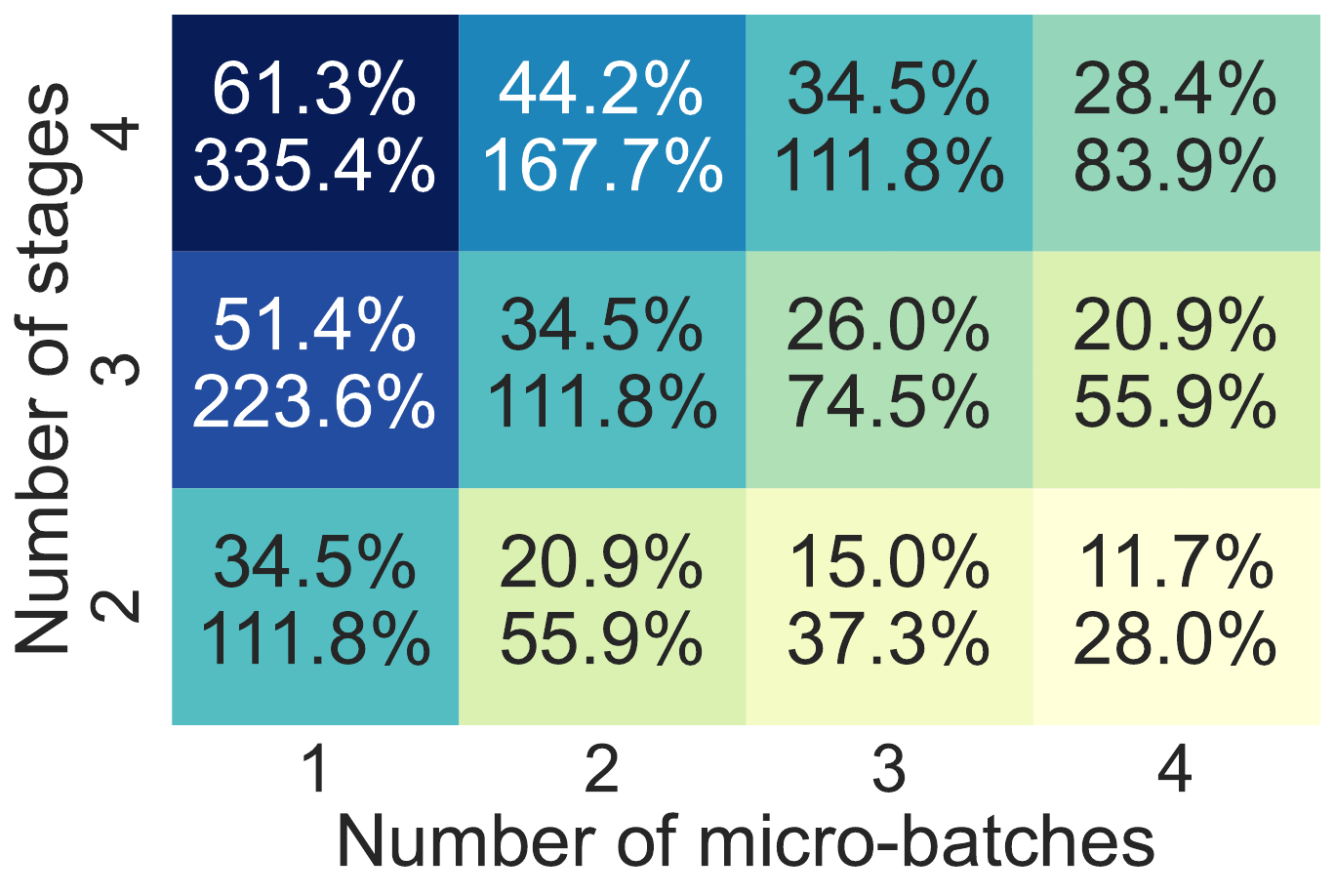}\label{fig:cldm_pipeline_bubble_ratio}}
    \caption{Ratio of pipeline bubble time to iteration time (upper) and ratio of pipeline bubble time to non-trainable part execution time (lower) at batch size 64 using FIFO-1F1B scheduling.}
    \label{fig:pipeline_bubble_ratio}
    \vspace{-20pt}
\end{figure}

We profile the iteration training time of two popular diffusion models (without self-conditioning) by pipelining their backbones under different model stage and micro-batch number settings, and executing the non-trainable part using data parallelism before backbone training.

Fig.~\ref{fig:pipeline_bubble_ratio} shows the pipeline bubble ratios, where the iteration time is the sum of pipeline training time of the backbone and the execution time of the non-trainable part in each training iteration.
Pipeline bubbles can take up to 68\% of the overall training time, which is quite significant, according to the upper line in Fig.~\ref{fig:pipeline_bubble_ratio}.
In the lower line, a ratio close to 1 indicates that the pipeline bubble time can be almost completely filled by scheduling the non-trainable part in pipeline bubbles, under the respective model stage and micro-batch numbers.
This observation motivates us to advocate pipeline bubble filling with the non-trainable part, and to study the detailed bubble filling strategies.

\begin{figure}[t]
    \centering
    \subfloat[Stable Diffusion v2.1]{\includegraphics[width=0.49\linewidth]{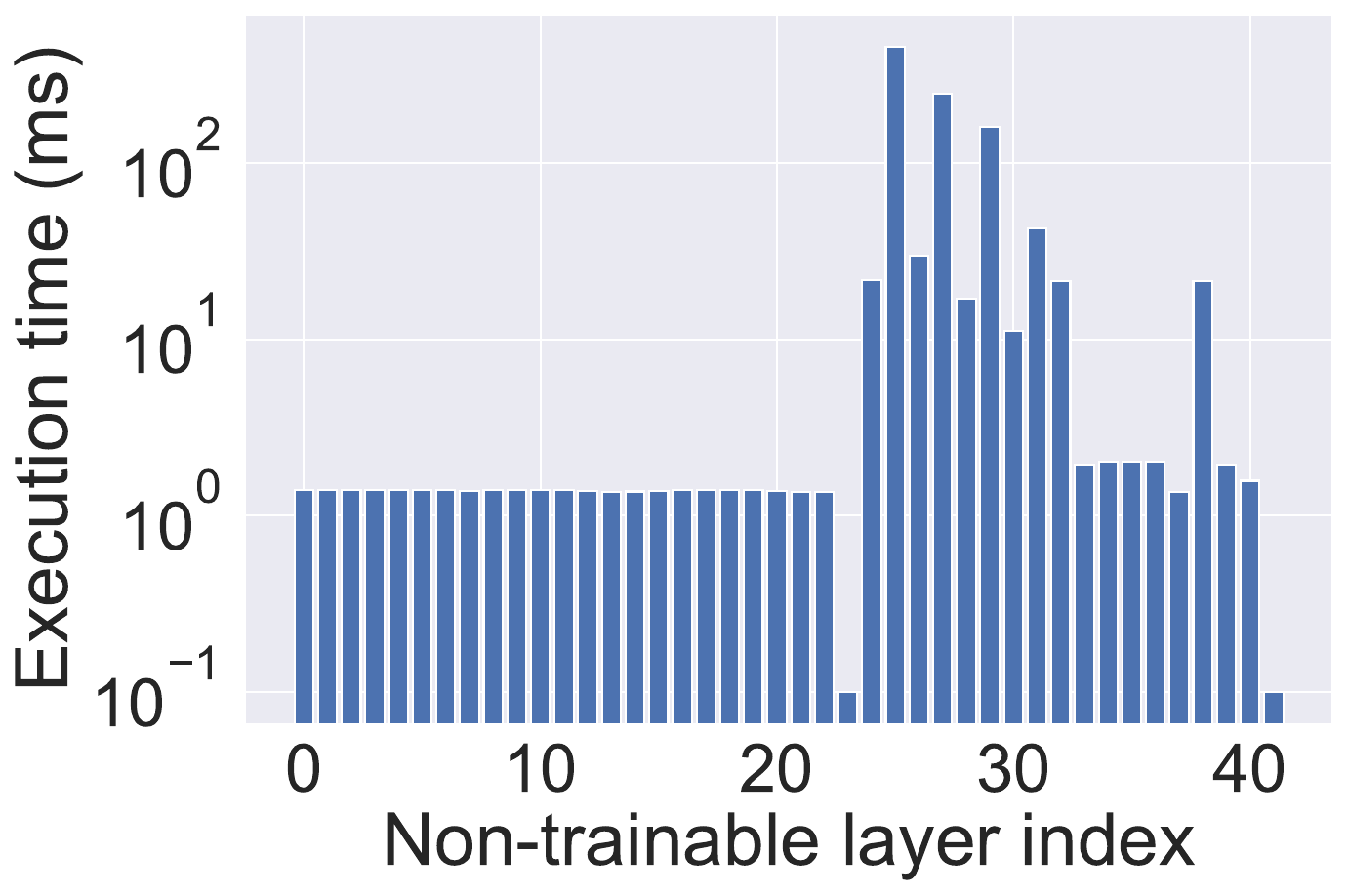}\label{fig:execution_times_sd}}\hskip1ex
    \subfloat[ControlNet v1.0]{\includegraphics[width=0.49\linewidth]{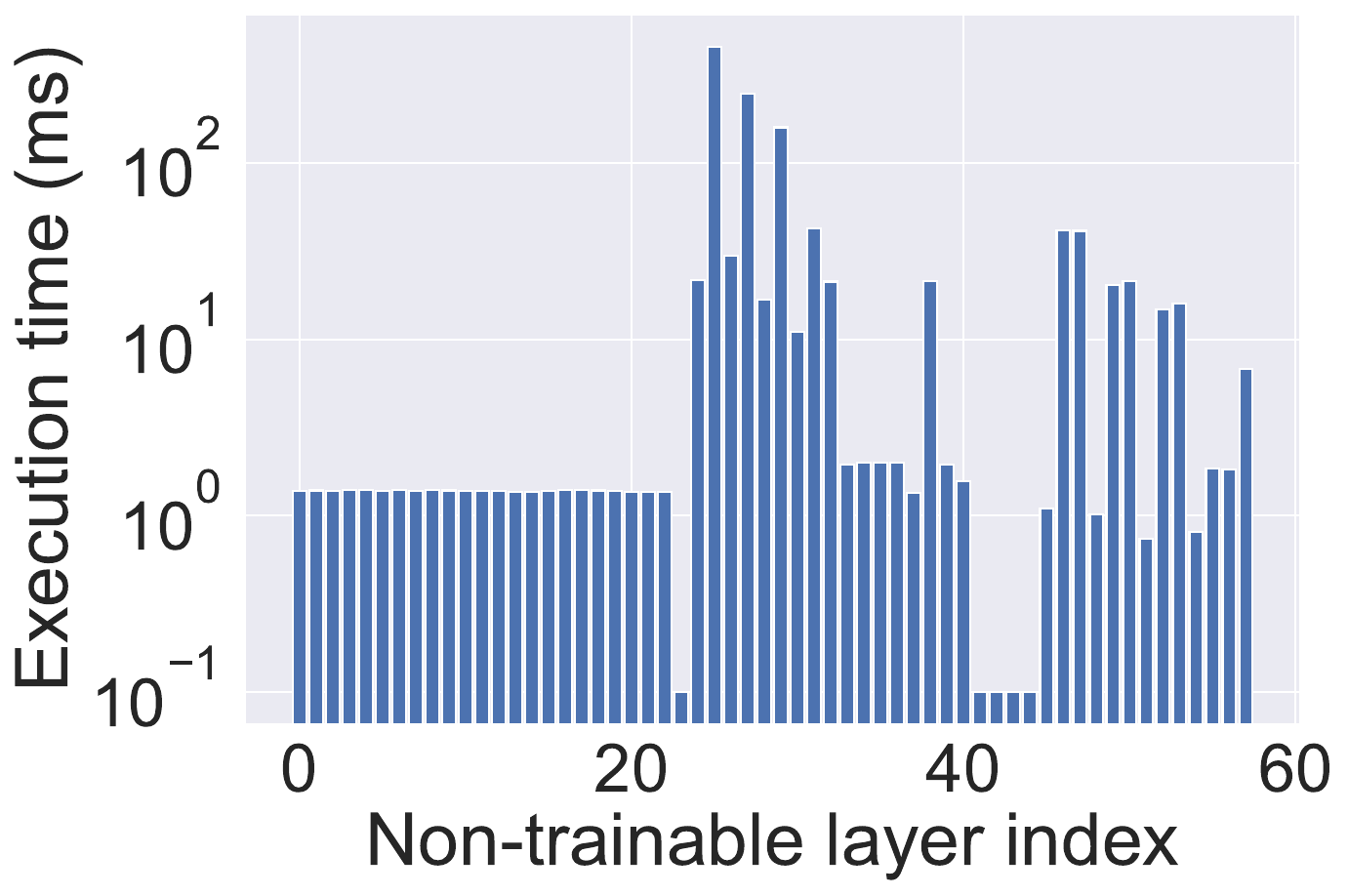}\label{fig:execution_times_cldm}}
    \caption{Execution time of non-trainable layers at batch size 64.}
    \label{fig:execution_times}
    \vspace{-15pt}
\end{figure}

Fig.~\ref{fig:execution_times} shows that many non-trainable layers (indexed 0 to 21) in both models have short execution times, which belong to the frozen text encoder.
Most layers (indexed 22 to 41) from the frozen image encoder take a moderate amount of time to compute (less than 30 ms).
Such a distribution of non-trainable layers with a large proportion of short and moderately long layer execution times provides excellent opportunities for executing individual layers in pipeline bubbles ranging from 10 to 100 ms.

\begin{figure}[t]
    \centering
    \subfloat[Stable Diffusion v2.1]{\includegraphics[width=0.49\linewidth]{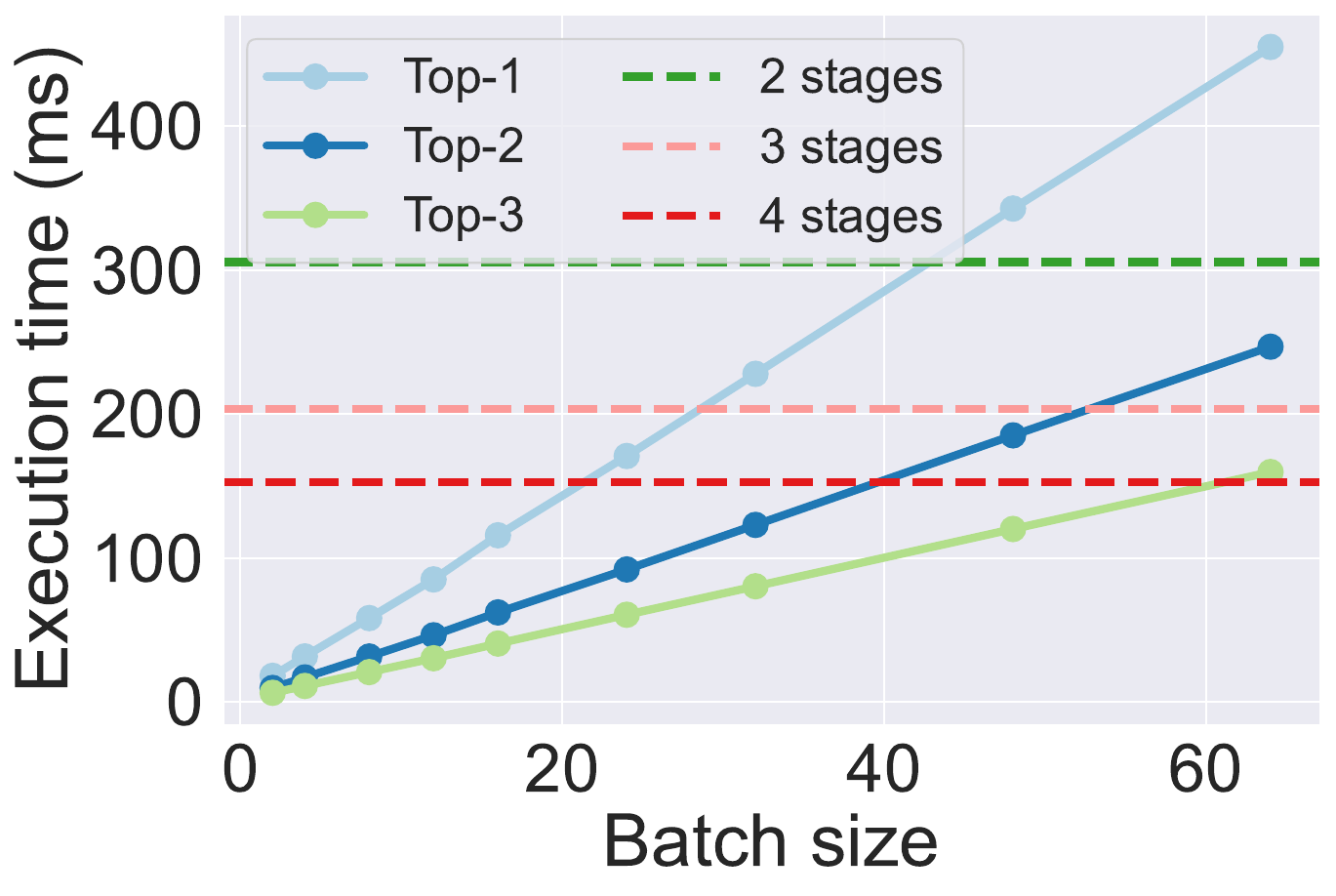}\label{fig:sd_partial_layer}}\hskip1ex
    \subfloat[ControlNet v1.0]{\includegraphics[width=0.49\linewidth]{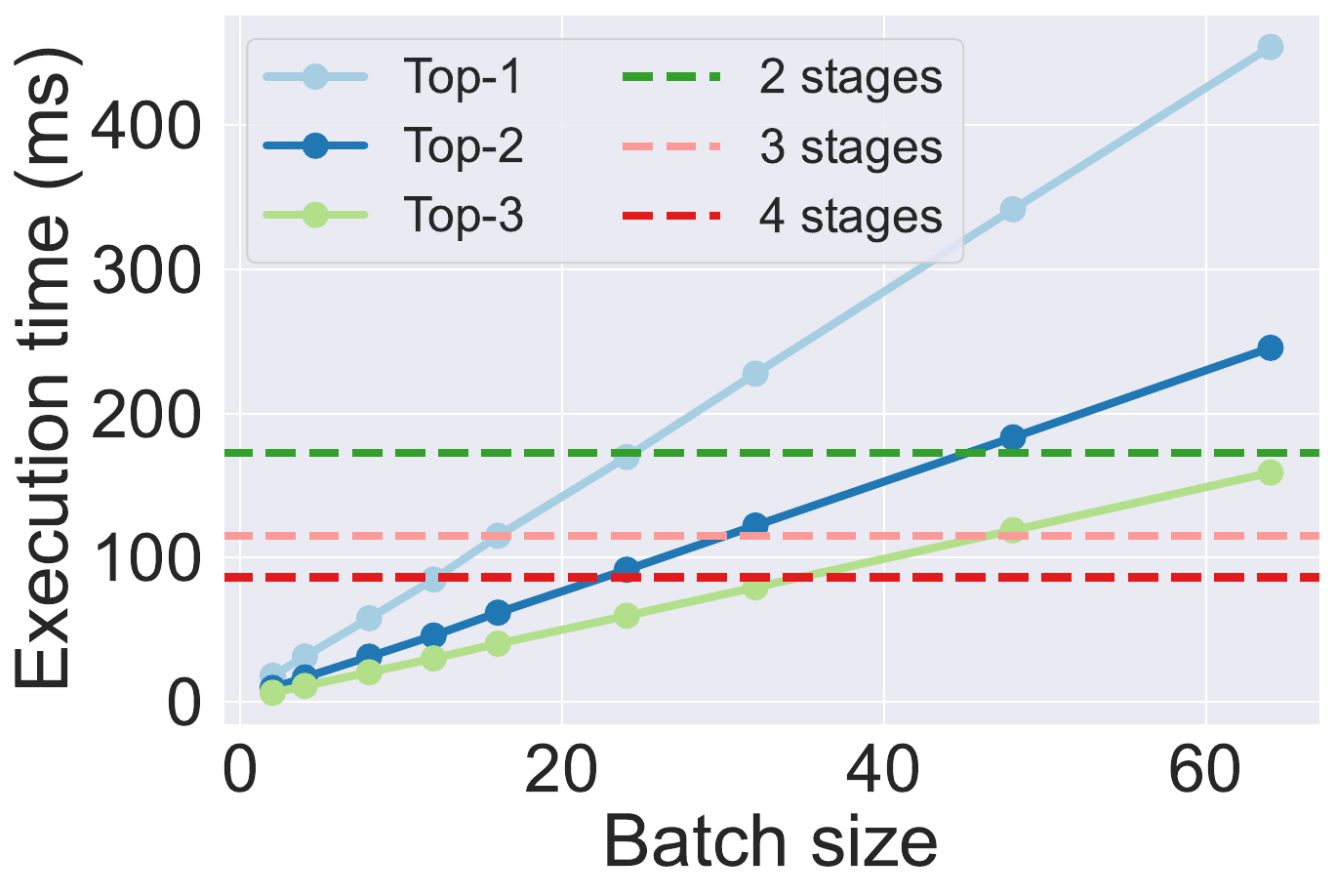}\label{fig:cldm_partial_layer}}
    \caption{Execution time of top-3 non-trainable layers with longest execution time under different batch sizes, compared to longest pipeline bubble time when there are 4 micro-batches and different number of stages at batch size 64 using FIFO-1F1B scheduling.}
    \label{fig:partial_layer}
    \vspace{-10pt}
\end{figure}

There are also some non-trainable layers with extra-long execution times (more than 400 ms), as shown in Fig.~\ref{fig:execution_times}. Such layers may not fit into any pipeline bubble.
Nevertheless, we observe that the layer execution time can be precisely controlled by adjusting the input batch size.
Fig.~\ref{fig:partial_layer} shows the execution times of the layers with the longest execution times at different batch sizes.
When the batch size is reduced to 16, most of these non-trainable layers can fit into the longest pipeline bubble obtained by the way we identify bubbles in Fig.~\ref{fig:fifo_ofob_schedule}, implying that we can run such layers in pipeline bubbles by partitioning their input.

We seek to design an efficient algorithm to schedule the execution of non-trainable layers into pipeline bubbles.

%% file: sources/system_overview.tex
\section{System design}\label{sec:system_design}
\begin{figure}[t]
    \centering
    \includegraphics[width=0.89\linewidth]{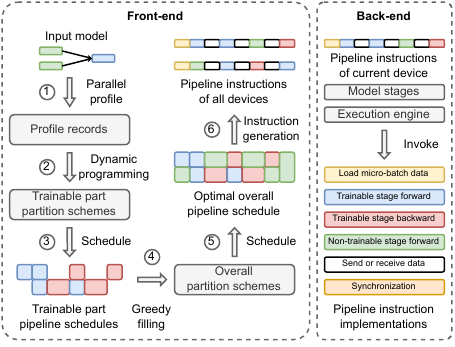}
    \vspace{-1.5em}
    \caption{The architecture of \name{}}
    \label{fig:system_overview}
    \vspace{-15pt}
\end{figure}

Fig.~\ref{fig:system_overview} presents an overview of~\name{}, which comprises of two modules:
(1) The front-end carries out our workflow of generating an optimized pipeline training schedule for an input diffusion model, including pipeline training configurations of the backbone(s) and bubble-filling strategies of the non-trainable part;
(2) The back-end is an execution engine that performs pipeline training according to the optimized pipeline schedule.

\subsection{Workflow}\label{sec:workflow}
\begin{figure}[t]
    \centering
    \includegraphics[width=0.85\linewidth]{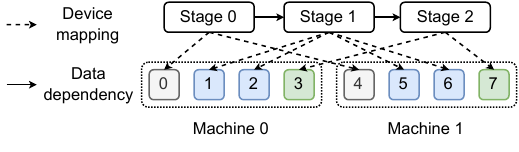}
    \vspace{-1em}
    \caption{\name{}'s data and pipeline parallelism. Devices in the same color run the same model stage.
    }\label{fig:partition_setting}
    \vspace{-1.5em}
\end{figure}

\name{} takes the diffusion model configuration, the training batch size, and the cluster configuration (i.e., number of machines and number of devices per machine) as inputs.
\name{} first performs parallel profiling on the entire cluster to obtain the model layer execution time at different batch sizes (step 1 in Fig.~\ref{fig:system_overview}), which is used in steps 2 to 5.
Based on the input specifications,~\name{} searches for pipeline training hyper-parameters as listed in Table~\ref{tab:hyperparameters}.
Note that~\name{} supports mixed pipeline and data parallelism, as shown in Fig.~\ref{fig:partition_setting}.
For each feasible hyper-parameter combination ($S$, $M$, and $D$),~\name{} generates a near-optimal partitioning scheme for the trainable backbone(s) (\S\ref{sec:trainable_partition}, step 2), including the number of layers in each model stage and the number of devices on which each stage replicates.
According to the corresponding pipeline schedule generated in step 3,~\name{} further partitions the non-trainable part and fills it into pipeline bubbles (\S\ref{sec:pipeline_bubble_filling}, step 4).
Then~\name{} generates the overall pipeline training schedules, and selects the optimal one with minimum iteration time (step 5).
Finally,~\name{} generates pipeline instructions for the back-end module according to the overall pipeline schedule (step 6).

\begin{table}[t]
\begin{center}
\begin{small}
\caption{Pipeline training hyper-parameters}
\begin{tabular}{|c|c|}
\hline
\textbf{Symbol} & \textbf{Description} \\ \hhline{|=|=|}
$S$ & Number of model stages \\ \hline
$M$ & \makecell{Number of micro-batches} \\ \hline
$D$ & \makecell{Pipeline parallel group size\tablefootnote{Pipeline parallel group is a minimum group of devices on which a complete set of pipeline communications is performed. In~\name{}, pipeline parallel group size (i.e., $D$) = world size (i.e., number of devices in the cluster) / data parallel degree.}}  \\ \hline
\end{tabular}
\label{tab:hyperparameters}
\vspace{-1.5em}
\end{small}
\end{center}
\end{table}

\subsection{Cross-iteration pipelining}\label{sec:cross_iteration_pipelining}
\begin{figure}[t]
    \centering
    \includegraphics[width=0.99\linewidth]{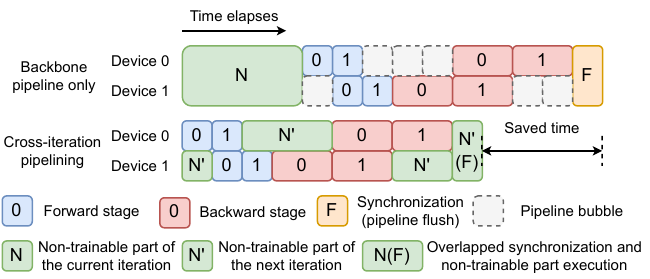}
    \vspace{-1em}
    \caption{Cross-iteration pipelining of a diffusion model. Numbers indicate the micro-batch index of a pipeline stage.
    }\label{fig:pipeline_bubble_filling}
    \vspace{-1em}
\end{figure}

For effective pipeline bubble filling that respects data dependencies between the non-trainable part and the backbone(s), \name{} advocates cross-iteration pipeline bubble filling, filling the bubble time of the backbone pipeline training in one iteration with the non-trainable part computation of the next iteration, as shown in Fig.~\ref{fig:pipeline_bubble_filling}.
Non-trainable layers can be computed in a data parallel manner without pipelining, following their inter-layer data dependencies.
At the end of a training iteration, the output of the non-trainable part is collected and divided into micro-batches according to the pipeline training configurations of the backbone(s).
In the next iteration, these intermediate results are loaded onto the correct devices and fed as input to the pipeline training of the backbone(s).
In addition, we only run the non-trainable part in the first iteration to enable such overlapping.
The cross-iteration pipeline is mathematically equivalent to data parallel and synchronous pipeline training.

%% file: sources/trainable.tex
\section{Backbone 
Partitioning}\label{sec:trainable_partition}
In this section, we present a unified dynamic programming approach to optimize partitioning and device assignment of the trainable part in diffusion models.

\input{sources/trainable_single}

\input{sources/trainable_cascaded}

\input{sources/trainable_self_conditioning}

%% file: sources/trainable_single.tex
\subsection{Single backbone}\label{sec:trainable_single}
We first consider a diffusion model with a single backbone.
The high-level idea is to analyze the critical path of FIFO-1F1B pipelining of the backbone and derive an upper bound on its execution time, to identify the optimal partitioning scheme that minimizes the execution time.
We use the notations in Table~\ref{tab:notations}.

\begin{table}[!h]
\begin{center}
\small
\caption{Notations}
\label{tab:notations}
\begin{tabular}{|c|c|}
\hline
\textbf{Symbol} & \textbf{Definition} \\ \hhline{|=|=|}
$L$ & Number of layers in backbone model\\ \hline
$\mathbf{B}$, $B$, $b$ & \makecell{Training batch size, micro-batch size \\ and number of samples in a partial-batch} \\ \hline
$\mathbf{S}, s$ & Set of model stages and model stage \\ \hline
\makecell{$\mathbf{P}^f_l(B)$, \\
$\mathbf{P}^b_l(B)$} & \makecell{Forward and backward computation time \\ of layer $l$ given batch size $B$} \\ \hline
\makecell{$\mathbf{C}_{l,l+1}^f(B)$, \\ $\mathbf{C}_{l+1,l}^b(B)$} & \makecell{Data size of communication in forward and \\ backward pass between layers $l$ \\ and $l+1$ given batch size $B$} \\ \hline
\makecell{$\mathbf{R}_{x}$, $\mathbf{L}_{x}$} & \makecell{Bandwidth and latency of communication type \\ $x$ (e.g., allreduce (ar), point-to-point (p2p))} \\ \hline
$\mathbf{G}_l(B)$ & Gradient size of layer $l$ given batch size $B$ \\ \hline
$\mathbf{O}_l(B)$ & Output size of layer $l$ given batch size $B$ \\ \hline
$T_S(s)$ & \makecell{Synchronization time of stage $s$} \\ \hline
$T_C(s)$ & \makecell{Compensation time of stage $s$} \\ \hline
$T_{0}$ & \makecell{Maximum micro-batch execution time \\ per stage or inter-stage communication time} \\ \hline
$T_{0}^{S-C}$ & \makecell{Maximum gap between synchronization time \\ and compensation time per stage} \\ \hline
$T_{B}$ & Length of a pipeline bubble (idle time) \\ \hline
\end{tabular}
\vspace{-5pt}
\end{center}
\end{table}

FIFO pipeline execution can be divided into 3 phases, i.e., warm-up, stable and cool-down, as shown in Fig.~\ref{fig:fifo_ofob_schedule}.
It launches micro-batch processing one by one in the warm-up phase and waits for all micro-batches to be completed in the cool-down phase.
When we enlarge the last stage's execution time to the longest among all stages, enforcing it on the critical path, the warm-up phase contains forward computation on $S - 1$ model stages (aka {\em forward stages}).
Similarly, the cool-down phase includes backward computation on $S - 1$ model stages (aka {\em backward stages}).
The stable phase of the critical path contains $M$ forward stages and $M$ backward stages, where $M$ is the number of micro-batches.
Therefore, there are a total of $2(M + S - 1)$ forward and backward stages on the critical path of the FIFO-1F1B pipeline schedule in total.
Considering the intermediate data communication between model stages in pipeline training, we add $S - 1$ inter-stage communications in the forward and backward passes, respectively, which then becomes $2(M + S - 1)+2(S - 1)$ forward and backward stages, together with communications on the critical path.

\begin{figure}[t]
    \centering
    \includegraphics[width=0.99\linewidth]{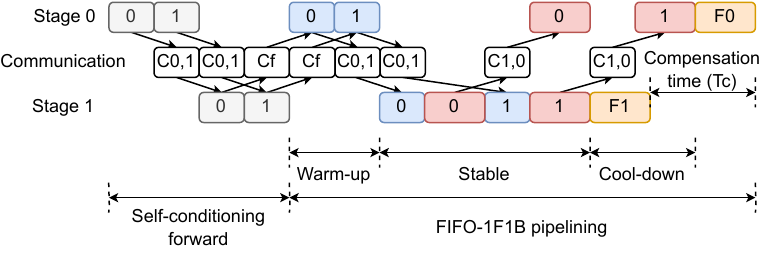}
    \vspace{-1.5em}
    \caption{FIFO-1F1B scheduling of pipelining a backbone model with 2 stages, 2 micro-batches and self-conditioning. The same color and number setting with Fig.~\ref{fig:pipeline_bubble_filling}. $C_{i,j}$ is communication from stage $i$ to stage $j$. $C_f$ feeds back the output of the backbone to stage 0. $F_i$ refers to the parameter synchronization of stage $i$, $T_c$ is the compensation time of stage 1.}
    \label{fig:fifo_scheduling}
    \vspace{-1.5em}
\end{figure}

We use $T_0$ to denote the maximum of the time to run the forward \textit{plus} backward computation of a micro-batch on a model stage, and the communication time between two stages, among all model stages.
Then we have an upper bound $T_0 (M + 2S - 2)$ on the execution time of the critical path.
We further consider the parameter synchronization time among the micro-batches and add $T^{S-C}_0$, i.e., the maximum gap between $T_{S}(s)$ and $T_{C}(s)$ to the pipeline training time of the backbone for all stages $s$, where $T_{S}(s)$ indicates the synchronization time of stage $s$ and $T_{C}(s)$ is used to compensate the overlapping time of parameter synchronization of stage $s$ and computation of later stages.
Fig.~\ref{fig:fifo_scheduling} gives an illustration.
Putting the above together, an upper bound on the FIFO-1F1B pipeline execution time is:

\begin{small}
\vspace{-1em}
\begin{equation}
    T^{max} = T_0 (M + 2S - 2) + T_{0}^{S-C}\label{eq:t_max_definition}
\end{equation}
\vspace{-2em}
\end{small}

We design a dynamic programming approach to identify the backbone partition and device assignment by minimizing $T^{max}$.
We order the $D$ devices in a pipeline parallel group into a chain according to their rank.
Let $W(L, S, r, D)$ denote $T_0$ when partitioning the first $L$ consecutive layers of the backbone into $S$ stages, with these stages placed on devices $1$ to $D$ and the last stage $s$ replicated on the last $r$ devices (of the $1$ to $D$ device chain).
Additionally, let $Y(L, S, r, D)$ denote $T_{0}^{S-C}$ under the same setting.
The optimal partition of the backbone into $S$ stages with the device placement of each stage can be computed by:

\begin{small}
\vspace{-2em}
\begin{equation}
    \min_{1\leq r\leq D}\{(M + 2S - 2)W(L,S,r,D) + Y(L,S,r,D)\}\label{eq:dp_opt_solution}
    \vspace{-5pt}
\end{equation}
\vspace{-1em}
\end{small}

$W(L, S, r, D)$ can be decomposed into sub-problems that further partition the first $l$ model layers into $S-1$ stages on the remaining $D-r$ devices, with the last stage replicated on $r'$ devices\footnotemark.
Then, $W(L, S, r, D)$ can be computed by the maximum of $W(l, S-1, r', D-r)$ and the estimation of $T_0$ by the last stage $s$ (i.e., $T_0(s)$), and $Y(L, S, r, D)$ can be computed in the same way, following Eqn.~(\ref{eq:t_0_definition}) to~(\ref{eq:dp_sub_problem}).
Then we add the range in Eqn.~(\ref{eq:sub_problem_range}) when optimizing Eqn.~(\ref{eq:dp_opt_solution}).
\footnotetext{{\small
Though we support different model stages using different data parallel degrees (e.g., $r \neq r'$), we find that such cases are rare. They can result in strange bubble filling schemes (\S\ref{sec:pipeline_bubble_filling}) and require complex implementations.
In evaluation (\S\ref{sec:evaluation}), we force all stages to have the same data parallel degree.
}}

\vspace{-2em}
\begin{small}
\begin{align}
T_0(s) = & \max\{\sum_{l< i \leq L} \mathbf{P}^f_i(\frac{B}{r}) + \sum_{l< i \leq L} \mathbf{P}^b_i(\frac{B}{r}), \notag \\
& \frac{\mathbf{C}^f_{l,l+1}(\frac{B}{r}) + \mathbf{C}^b_{l+1,l}(\frac{B}{r})}{\mathbf{R}_{p2p}} + 2\mathbf{L}_{p2p} \}\label{eq:t_0_definition} \\
T_S(s) = & \sum_{l< i \leq L}\mathbf{G}_{i}(\frac{B}{r}) / \mathbf{R}_{ar} + \mathbf{L}_{ar} \label{eq:t_s_definition} \\
T_{C}(s) = & \sum_{l< i \leq L} \mathbf{P}^b_i(\frac{B}{r}) \label{eq:c_definition} \\
T_{0}^{S-C}(s) = & T_S(s) - T_C(s) \\
W(L, S, r, D) = & \max\{W(l,S-1,r',D-r), T_0(s)\} \\
Y(L, S, r, D) = & \max\{Y(l,S-1,r',D-r), T_{0}^{S-C}(s)\} \label{eq:dp_sub_problem}\\
\forall l, r', \phantom{=} & 1\leq l\leq L-1,\ 1\leq r'\leq D-r \label{eq:sub_problem_range}
\end{align}
\end{small}
\vspace{-2em}

Here $B$ is the micro-batch size, $\mathbf{P}^{\{f/b\}}_i(\frac{B}{r})$ is the forward / backward computation time of layer $i$ given local batch size $\frac{B}{r}$.
$\mathbf{C}_{l,l+1}^f(\frac{B}{r})$ and $\mathbf{C}_{l+1,l}^b(\frac{B}{r})$ are data sizes in forward and backward pass between layer $l$ and $l+1$ given local batch size $\frac{B}{r}$.
$\mathbf{G}_i(\frac{B}{r})$ is the gradient size of layer $i$ given local batch size $\frac{B}{r}$.
$\mathbf{R}_x$ and $\mathbf{L}_x$ are bandwidth and latency of communication type $x$, while $ar$ indicates all-reduce used in synchronization and $p2p$ indicates point-to-point communication between model stages.

Note that a sub-problem in Eqn.~(\ref{eq:dp_sub_problem}) does not know the partition scheme of all subsequent layers (with indices greater than $l$) for computing the compensation time $T_C$.
Instead, we use a lower bound of $T_C$ in Eqn.~(\ref{eq:c_definition}), i.e., the sum of the backward computation time of all these layers on $r$ devices.

%% file: sources/trainable_cascaded.tex
\subsection{Multiple backbones}\label{sec:trainable_cascaded}
For a cascaded diffusion model with multiple backbones, 
we advocate bidirectional pipelining to train the backbones on the same set of devices (instead of using a separate set of devices to train each backbone), in order to utilize the devices more efficiently.
In particular, we leverage bidirectional pipelining~\cite{li2021chimera} to train multiple backbones, with each backbone pipelining in different direction.
Here, we consider pipelining from low-rank device to high-rank device as the down direction and vice versa, and the corresponding pipelines are down and up pipelines.

Consider 2 backbones in a CDM. As shown in Fig.~\ref{fig:bidirectional_schedule}, the duration of the stable phase of the critical path in bidirectional pipelining differs from unidirectional pipelining while the duration of the warm-up and cool-down phases are not affected.
We calculate the number of paired forward and backward stages between the down and up pipelines ($M_{CDM}$), and derive an upper bound on bi-directional pipeline execution time for training two backbones:

\vspace{-2em}
\begin{small}
\begin{align}
    T_{0,CDM} &= \max\{T_{0,down}, T_{0,up} \} \label{eq:t_0_cdm_bi_directionaL_definition}\\
    T^{S-C}_{0,CDM} &= \max\{T^{S-C}_{0,down}, T^{S-C}_{0,up} \}\label{eq:t_0_gap_cdm_bi_directionaL_definition}\\
    T^{max}_{CDM} &= (M_{CDM} + 2S - 2) T_{0,CDM} + T^{S-C}_{0,CDM}\label{eq:t_max_fifo_cdm_definition}
\end{align}
\end{small}
\vspace{-2em}

Here $T_{0,\{down/up\}}$ is the maximum of the time to perform the forward and backward computation of a micro-batch and the communication time between two stages in the down or up pipeline among all model stages.
The sub-problem in the dynamic programming approach in bi-directional pipelining should decide partitioning and placement of model stages for both backbones.
Let $W(L_d, L_u, S, r, D)$ denote $T_{0,CDM}$ when partitioning the last $L_d$ and the first $L_u$ consecutive layers of the down- and up-pipelined backbones, respectively, into $S$ stages, while placing them on $D$ devices and replicating the last stage $s_{d}$ and the first stage $s_{u}$ of the two backbones on the last $r$ devices of the $1$ to $D$ device chain, and we have $Y(L_d, L_u, S, r, D)$ similarly.
The optimal partitioning of two backbones can be computed by:

\vspace{-2em}
\begin{small}
\begin{align}
\min_{1\leq r\leq D} & \{(M_{CDM} + 2S - 2)W(L_d,L_u,S,r,D) \notag \\
& \phantom{\{} + Y(L_d,L_u,S,r,D)\}\label{eq:dp_opt_solution_cdm}
\end{align}
\end{small}
\vspace{-2em}

In Eqn.~(\ref{eq:w_cdm_definition}) and~(\ref{eq:y_cdm_definition}) we give the definition of $W(L_d, L_u, S, r, D)$ and $Y(L_d, L_u, S, r, D)$.
Eqn.~(\ref{eq:sub_problem_range_cdm}) presents the additional optimization range.
Communication in the bidirectional pipelines may compete for resources, and we reasonably enlarge the communication time in Eqn.~(\ref{eq:t_0_definition}) by a factor of 2 (as there are two pipelining directions).
Other equations are the same with~\S\ref{sec:trainable_single}.

\vspace{-2em}
\begin{small}
\begin{align}
W(L_d, L_u, S, r, D) = & \max\{ W(l_d, l_u, S - 1, r', D - r), \notag \\
& T_{0}(s_d), T_{0}(s_u)\} \label{eq:w_cdm_definition}\\
Y(L_d, L_u, S, r, D) = & \max\{ Y(l_d, l_u, S - 1, r', D - r), \notag \\
& T^{S-C}_{0}(s_d), T^{S-C}_{0}(s_u)\} \label{eq:y_cdm_definition}\\
\forall\ l_{d/u}, r', \phantom{=} & 1\leq l_{\{d/u\}} \leq L_{\{d/u\}} - 1, \notag \\
& 1 \leq r' \leq D - r \label{eq:sub_problem_range_cdm}
\end{align}
\end{small}
\vspace{-2em}

For a diffusion model with more than two backbones, we can divide the backbones into two groups, one to be pipelined in each direction. We then combine stages of the backbones in the same pipeline direction to form a larger model stage and apply our design for bi-directional pipelining accordingly.

%% file: sources/trainable_self_conditioning.tex
\subsection{Backbone(s) with self-conditioning}\label{sec:trainable_self_conditioning}
\name{} performs self-conditioning on the same device to eliminate unnecessary parameter storage and updating, as shown in Fig.~\ref{fig:fifo_scheduling}.
There is an additional forward pass at each model stage, and Eqn.~(\ref{eq:t_0_definition}) is changed to:

\vspace{-2em}
\begin{small}
\begin{align}
T_{0,SC}(s) = & \max\{2\sum_{l< i \leq L} \mathbf{P}^f_i(\frac{B}{r}) + \sum_{l< i \leq L} \mathbf{P}^b_i(\frac{B}{r}), \notag \\
& \frac{2\mathbf{C}^f_{l,l+1}(\frac{B}{r}) + \mathbf{C}^b_{l+1,l}(\frac{B}{r})}{\mathbf{R}_{p2p}} + 3\mathbf{L}_{p2p} \label{eq:t_sc_0_definition}
\end{align}
\end{small}
\vspace{-2em}

In addition, the communication time for sending the output from the last stage to the first stage ($C_{f}$ in Fig.~\ref{fig:fifo_scheduling}) should be considered.
We use a point-to-point transmission time as the upper bound of this feedback time: $T_{F} = \mathbf{O}_{L}(B) / \mathbf{R}_{p2p} + \mathbf{L}_{p2p}$, where $\mathbf{O}_{L}(\frac{B}{r})$ is the output size of the last layer $L$ at local batch size $\frac{B}{r}$.
The upper bound on the pipeline execution time with self-conditioning is:

\vspace{-1.5em}
\begin{equation}\small
    T^{max}_{SC} = (M + 2S - 2)T_{0,SC} + T^{S-C}_{0} + T_F\label{eq:t_max_fifo_sc_0_definition}
\end{equation}
\vspace{-2em}

The dynamic programming formulation remains the same as in~\S\ref{sec:trainable_single}. Since self-conditioning is usually randomly activated during the training process with a certain probability $p$ (0.5 in~\cite{chen2022analog}), the formulation optimizes an expectation of $T^{max}_{SC}$ and $T^{max}$.

In case self-conditioning is applied to CDMs, we can readily extend the formulation in~\S\ref{sec:trainable_cascaded} by counting the additional number of forward stages in the critical path.

%% file: sources/pipeline_bubble_filling.tex
\section{Pipeline bubble filling}\label{sec:pipeline_bubble_filling}
In~\name{}, we divide the pipeline idle time along the timeline and define a pipeline bubble using a tuple (\textit{start time}, \textit{end time}, \textit{idle devices}) so that a bubble contains the same number of idle devices in its time span. 
For example, in Fig.~\ref{fig:fifo_ofob_schedule}, the first pipeline bubble is in the first time slot with idle devices 1 to 3.
Pipeline bubble filling is always performed under the cross-iteration style of pipelining~(\S\ref{sec:cross_iteration_pipelining}), regardless of the number of backbones and whether self-conditioning is applied.
We also define a partial-batch layer using a tuple (\textit{component index}, \textit{layer index}, \textit{number of samples in partial-batch}).

Non-trainable components in a diffusion model may have inter-dependencies (e.g., ControlNet~\cite{zhang2023adding}), and layers within each component are linearly dependent.
We schedule the execution of non-trainable components in pipeline bubbles following a topological order of components according to their dependencies.
Especially, we fill in the pipeline bubbles sequentially in their chronological order\footnotemark.
To fill a pipeline bubble, we consider all of the components that are ready at the time, i.e., their dependencies are resolved.
Whenever a component becomes ready, we add it to the set of ready components.
\footnotetext{{\small
For bubble filling efficiency, we only identify pipeline bubbles longer than 10 ms, which is empirically greater than the cost of setting up inputs and outputs for pipeline bubble filling.
Chronological order of pipeline bubbles is achieved by analyzing the pipeline schedule, which is simulated using the profiled results obtained in step 1 of Fig.~\ref{fig:system_overview}.
All proposed algorithms work offline only.
}}

\begin{figure}[!t]
    \centering
    \includegraphics[width=0.99\linewidth]{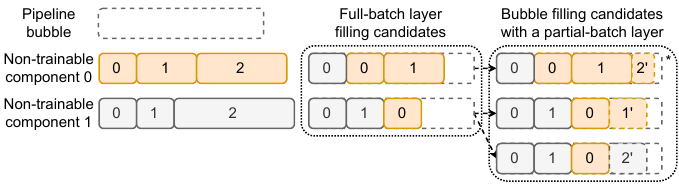}
    \vspace{-2mm}
    \caption{Full-batch bubble filling candidates and corresponding candidates with a partial-batch layer. Numbers indicate the index of the non-trainable layer. 1' and 2' denote partial-batch layers. * marks the candidate with the longest execution time.}
    \label{fig:partition_candidates}
    \vspace{-15pt}
\end{figure}

An efficient algorithm is designed to fill a pipeline bubble with ready components (Alg.~\ref{alg:full_partition_candidates}).
Its input mainly includes the bubble time $T_B$ and the number of idle devices $d$, a list $u$ containing the index of the starting layer in each currently ready component (layers of a component can be executed in multiple bubbles).
It first finds candidates $K$ containing full-batch layers of ready components to fill the current pipeline bubble (Alg.~\ref{alg:initial_partition_candidates}), whose execution completes within the bubble time.
Then it adds at most one layer from a component to be executed on a partial batch in the remaining bubble time to each candidate, and finally it produces the optimal bubble filling scheme with the longest execution time (not exceeding the bubble time), as shown in Fig.~\ref{fig:partition_candidates}.
Note that the component layers assigned to the bubble are executed in a data parallel manner at local batch size $\frac{\mathbf{B}}{d}$.

Input of Alg.~\ref{alg:initial_partition_candidates} includes the input of Alg.~\ref{alg:full_partition_candidates}, and the component index $i$ that it focuses on.
It finds bubble filling candidates containing full-batch layers in a recursive manner: assuming layers from components with indices smaller than $i$ are already considered, it adds layers from component $i$ to the candidate.
Alg.~\ref{alg:initial_partition_candidates} first computes how many layers can be added at most from line~\ref{lino:ipc_cumulative_start} to~\ref{lino:ipc_cumulative_end}, where $\mathbf{P}_{i,u_i + k}^f(\frac{\mathbf{B}}{d})$ is the computation time of layer $u_i + k$ of component $i$ given local batch size $\frac{\mathbf{B}}{d}$.
Then it adds different numbers of layers to the candidate (with total execution time not exceeding $T_B$), and recursively calls itself to add layers from the next component $i + 1$ from line~\ref{lino:ipc_sub_problem_start} to~\ref{lino:ipc_sub_problem_end}.
Alg.~\ref{alg:initial_partition_candidates} returns a list $K$ containing bubble filling candidates, where each candidate is a list containing $n$ elements ($n$ is the number of ready components), with each element containing the indices of the layers of that component to be executed in the bubble.

\begin{algorithm}[!t]\small
\caption{Filling One Pipeline Bubble}\label{alg:full_partition_candidates}
\begin{algorithmic}[1]
\REQUIRE Number of ready non-trainable components $n$, training batch size $\mathbf{B}$, pipeline bubble time $T_B$, number of idle devices $d$, indices of starting layers of components $u$ (list, length is $n$), numbers of layers of components $\mathbf{L}$ (list, length is $n$)
\ENSURE Optimal bubble filling candidate $k^*$
\STATE $K_0, K \gets $ emptyList(), \textbf{FFC}($n$, $\mathbf{B}$, $T_B$, $d$, $u$, $\mathbf{L}$, $0$)
\FOR{$k$ in $K$, $h$ in $0,\dots,n - 1$\label{lino:fpc_iterate_start}}
    \STATE $b \gets $ maximum of getValidNumSamples($\mathbf{B}$, $d$), s.t., \footnotemark\label{lino:get_valid_ns}\label{lino:fpc_add_partial_start}
    \STATE \phantom{$b \gets $} $T_B \geq \sum_{i \in [n], j \in [k_i]} \mathbf{P}_{i,u_i + j}^f(\frac{\mathbf{B}}{d}) + \mathbf{P}_{h,u_h + k_h}^f(\frac{b}{d})$ \COMMENT{Bubble time should be greater than the sum of execution time of candidate $k$ and a partial-batch layer} \label{lino:fpc_add_partial_end}
    \STATE $K_0$.append(($k$, ($h$, $u_h + k_h$, $b$)) \COMMENT{Add candidate $k$ enhanced with a partial-batch layer ($h$, $u_h + k_h$, $b$)} \label{lino:fpc_add_partial}
\ENDFOR
\STATE \textbf{return} the candidate in $K_0$ with the longest execution time
\end{algorithmic}
\end{algorithm}
\footnotetext{{\small Here $[n] := \{0, 1, \dots, n-1\},\ [k_i] := \{0, 1, \dots, k_i - 1\}$.}}

\begin{algorithm}[t]\small
\caption{\textbf{FFC} - \textbf{F}ull-batch Layer Bubble \textbf{F}illing \textbf{C}andidates}\label{alg:initial_partition_candidates}
\begin{algorithmic}[1]
\REQUIRE $n$, $\mathbf{B}$, $T_B$, $d$, $u$, $\mathbf{L}$, current component index $i$
\ENSURE bubble filling candidates $K$
\STATE $t, k_0, K \gets 0, 0,$ emptyList()
\WHILE{$t + \mathbf{P}_{i,u_i + k_0}^f(\frac{\mathbf{B}}{d}) \leq T_B$ and $u_i + k_0 < \mathbf{L}_i$\label{lino:ipc_cumulative_start}}
    \STATE $t \gets t + \mathbf{P}_{i,u_i + k_0}^f(\frac{\mathbf{B}}{d})$ \COMMENT{Cumulative execution time} \label{lino:compute_layer_time}
    \STATE $k_0 \gets k_0 + 1$
\ENDWHILE\label{lino:ipc_cumulative_end}
\IF{$i = n - 1$}
    \STATE \textbf{return} [[$k_0$]] \COMMENT{Add all $k_0$ layers to the candidate as it is the last component}
\ELSE
    \FOR{$k$ in $k_0,\dots,0$\label{lino:ipc_sub_problem_start}}
        \STATE $T_{B}' \gets T_B - \sum_{h\in [k]}\mathbf{P}_{i,u_i+h}^f(\frac{\mathbf{B}}{d})$ \COMMENT{Remaining bubble time after adding $k$ layers to the candidate}
        \STATE $K' \gets$ \textbf{FFC}($n$, $\mathbf{B}$, $T_{B}'$, $d$, $u$, $\mathbf{L}$, $i + 1$) \label{lino:ipc_sub_problem}
        \STATE $K$.extend([concat([$k$], $k'$) for $k'$ in $K'$])
    \ENDFOR\label{lino:ipc_sub_problem_end}
    \STATE \textbf{return} $K$
\ENDIF
\end{algorithmic}
\end{algorithm}

Then for each bubble filling candidate $k$ in $K$, an additional partial-batch layer is added to it.
Especially, we identify a layer that is the subsequent layer following the scheduled full-batch layers in the candidate, as well as a partial batch to process, whose execution can occupy the longest of the remaining bubble time (line~\ref{lino:fpc_add_partial_start} to~\ref{lino:fpc_add_partial_end} of Alg.~\ref{alg:full_partition_candidates}).
We then choose among the enhanced bubble filling candidates with a partial-batch layer each, the one achieving the longest execution time to maximally utilize the idle time.

\begin{figure}[t]
    \centering
    \includegraphics[width=0.85\linewidth]{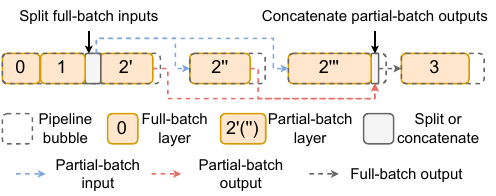}
    \caption{Input split and output concatenation of partial-batch layer's processing among pipeline bubbles. The partial-batch layer 2 of a non-trainable component is scheduled in 3 consecutive pipeline bubbles.}
    \label{fig:partial_mini_batch_processing}
    \vspace{-15pt}
\end{figure}

To decide the partial batch for the extra layer to process in a pipeline bubble (function getValidNumSamples in line~\ref{lino:get_valid_ns} of Alg.~\ref{alg:full_partition_candidates}), 
we follow two principles:
(1) The local batch size $b / d$ should not be too small, as otherwise the benefit of inserting a partial-batch layer will not compensate for the overhead of handling its input and output, as illustrated in Fig.~\ref{fig:partial_mini_batch_processing};
(2) $b / d$ should be a \textit{regular} value to avoid potential kernel performance degradation at unusual batch sizes.
We empirically use 4, 8, 12, 16, 24, 32, 48, 64 and 96 as the local batch size candidates.
If pipeline bubbles cannot completely accommodate the non-trainable part, the remaining part will be executed after pipelining completes.

Furthermore, after introducing a partial-batch layer ($h$, $u_h + k_h$, $b$) in a pipeline bubble (line~\ref{lino:fpc_add_partial} of Alg.~\ref{alg:full_partition_candidates}), the layer $u_h + k_h$ of component $h$ is the first ready layer of that component to be considered when filling the following pipeline bubbles, and it is treated as a full-batch layer on the remaining batch. 
In this way, this layer can be scheduled to process all or part of the remaining batch in a subsequent pipeline bubble.
Fig~\ref{fig:partial_mini_batch_processing} shows an example of scheduling part of the remaining batch in a subsequent (i.e., the second) pipeline bubble.

%% file: sources/evaluation.tex
\section{Evaluation}\label{sec:evaluation}
We build~\name{} on PyTorch 2.0.1 and CUDA 11.7 with 20k LoC in Python, and integrate it with DeepSpeed 0.8.3 to support pipeline and data parallel training. Communication operations are implemented using PyTorch's distributed communication package and NCCL 2.17.1.
Though~\name{} is integrated into DeepSpeed, it is easy to migrate~\name{} to other vendor frameworks.
We only need to switch to the new launching method and replace the communication and optimizer implementations with corresponding implementations.

\noindent \textbf{Test-bed}
We conduct our experiments on a cluster of 8 Amazon EC2 p4de.24xlarge machines, each containing 8 NVIDIA A100-80GB GPUs and 96 vCPU cores.
The inter-node connection (EFA) bandwidth is 400 Gbps and the intra-node connection (NVSwitch) bandwidth is 600 GBps.

\noindent \textbf{Models} We train these models: Stable-Diffusion~\cite{rombach2022high} v2.1, ControlNet~\cite{zhang2023adding} v1.0, CDM-LSUN and CDM-ImageNet~\cite{ho2022cascaded}.
For CDM-LSUN, we train its 2 backbones using bi-directional pipelining.
For CDM-ImageNet, we only train its second and third backbones because training all of them will exceed the GPU memory.
The backbones of the same CDM are trained under the same batch size.
The input configurations of all models (Table~\ref{tab:model_configurations}) are the same as in their original papers.

\begin{table}[ht]
\vspace{-2.5em}
\begin{center}
\begin{small}
\caption{Diffusion models and training configurations}
\begin{tabular}{|c|c|c|}
\hline
\textbf{Model} & \textbf{Input shape} & \textbf{Self-cond}\\ \hhline{|=|=|=|}
Stable Diffusion v2.1 & \multirow{2}{*}{512x512} & \multirow{2}{*}{Enabled} \\ \cline{1-1}
ControlNet v1.0 & & \\ \hline
CDM-LSUN & \multirow{2}{*}{\makecell{64x64, 128x128 \\ (2 image inputs)}} & \multirow{2}{*}{Not enabled} \\ \cline{1-1}
CDM-ImageNet & & \\ \hline
\end{tabular}
\label{tab:model_configurations}
\end{small}
\end{center}
\vspace{-2em}
\end{table}

\noindent \textbf{Baselines} We run DeepSpeed~\cite{rasley2020deepspeed} with vanilla distributed data parallelism (DDP) and ZeRO-3~\cite{rajbhandari2021zero} as baselines for data parallel training.
We use GPipe~\cite{huang2019gpipe} and SPP~\cite{luo2022efficient} as baselines of pipeline parallelism, which perform the backbone only pipelining in Fig.~\ref{fig:pipeline_bubble_filling}.
For GPipe that partitions a model into stages with equal number of layers, we evaluate it with 2 pipeline stages and 4 micro-batches.
For SPP that solves a dynamic programming problem to optimize model partitioning, we perform the same hyper-parameter searching as in~\name{}.
When self-conditioning is enabled, we also run the extra-forward part in the way shown in Fig.~\ref{fig:fifo_scheduling} for pipeline parallel baselines.
Bubble filling is not performed for pipeline parallel baselines.

For cascaded diffusion models, data parallel training is performed in two ways:
(1) Training multiple backbones in sequential using all devices, i.e., DeepSpeed(-ZeRO-3)-S;
(2) Training multiple backbones in parallel on evenly partitioned sets of devices, i.e., DeepSpeed(-ZeRO-3)-P, which is the default strategy in many CDM works.
Both SPP and GPipe do not apply to CDM, because they do not support pipelining of multiple models.

\noindent \textbf{Metrics} We present the training throughput in terms of the number of samples processed per second.
The throughput of DeepSpeed(-ZeRO-3)-S and DeepSpeed(-ZeRO-3)-P is computed by $\small \frac{\mbox{total batch size of all backbones}}{\mbox{sum of iteration time of all backbones}}$ and sum of $\small \frac{\mbox{batch size}}{\mbox{iteration time}}$ of all backbones, respectively.
We also present the pipeline bubble ratio of~\name{} and pipelined baselines, which is computed by $\frac{\sum_{b\in pipeline\ bubbles} T_b \times d_b}{iteration\_time \times total\_num\_devices}$, where $T_b$ and $d_b$ are the duration and the number of idle devices of the bubble $b$.

\subsection{Training throughput}
\begin{figure*}[!t]
    \centering
    \subfloat[Stable Diffusion v2.1]{\includegraphics[width=0.99\linewidth]{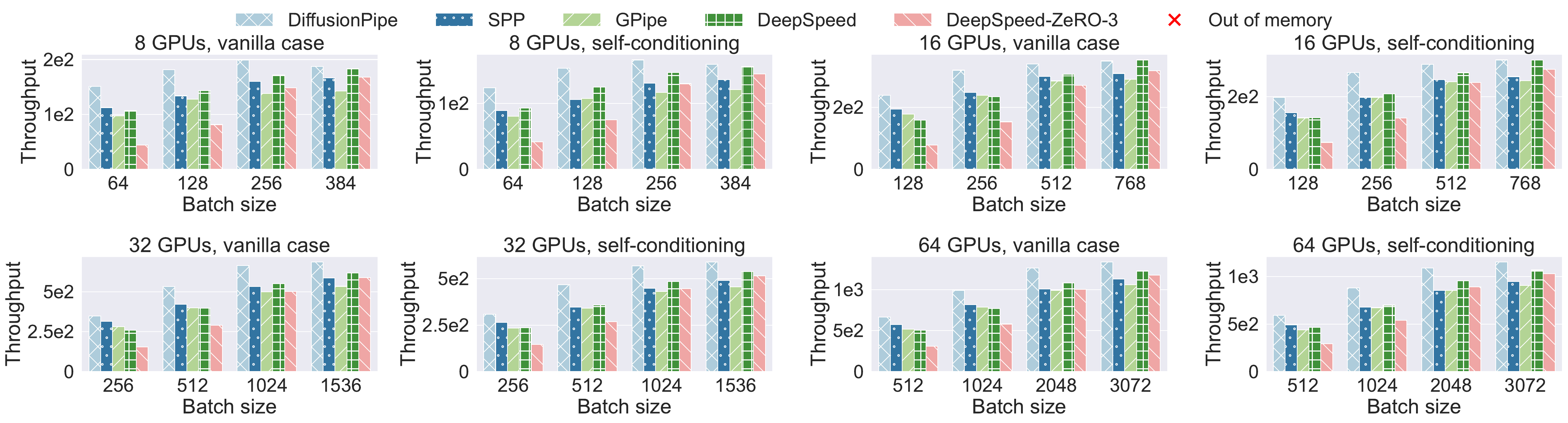}\label{fig:throughput_sd_v21}}\\
    \vspace{-1em}
    \subfloat[ControlNet v1.0]{\includegraphics[width=0.99\linewidth]{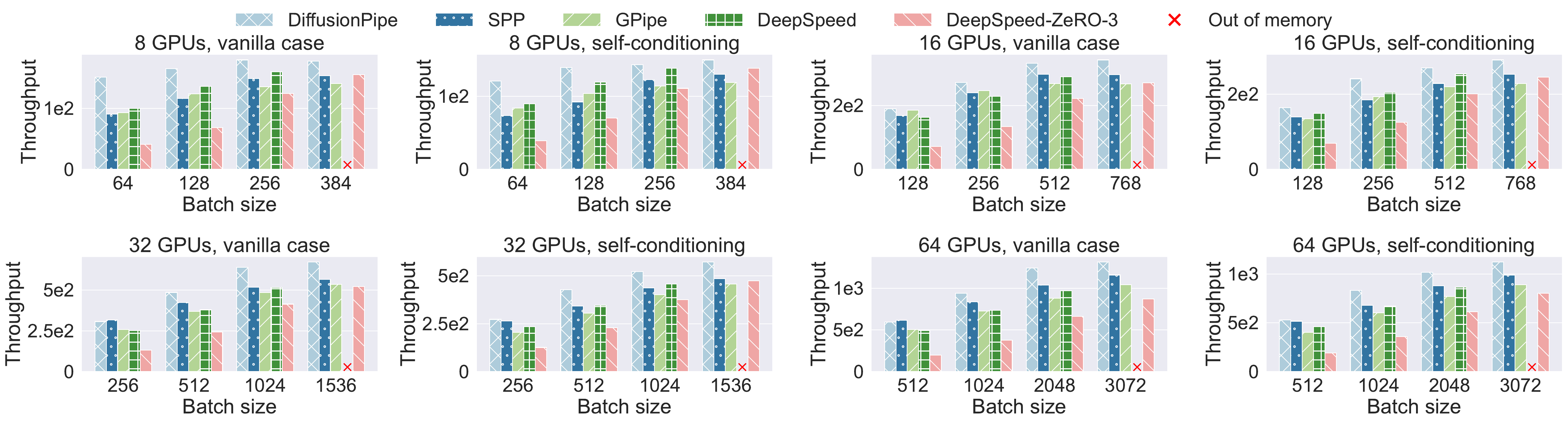}\label{fig:throughput_cldm_v10}}\\
    \vspace{-1em}
    \subfloat[CDM-LSUN]{\includegraphics[width=0.99\linewidth]{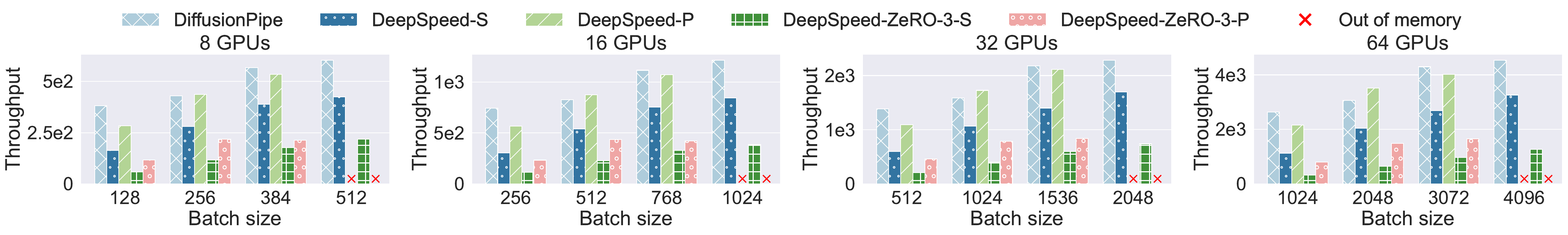}\label{fig:throughput_cdm_lsun}}\\
    \vspace{-1em}
    \subfloat[CDM-ImageNet]{\includegraphics[width=0.99\linewidth]{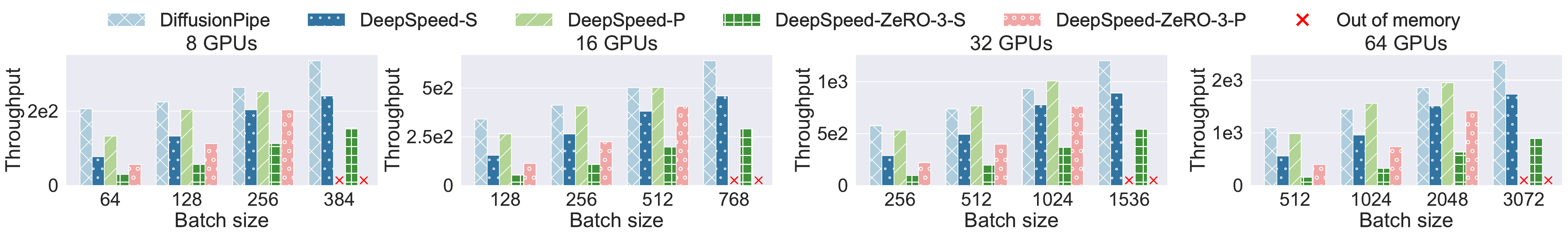}\label{fig:throughput_cdm_imagenet}}
    \vspace{-0.5em}
    \caption{Training throughput (samples/second)}
    \label{fig:throughput}
    \vspace{-1.5em}
\end{figure*}

In Fig.~\ref{fig:throughput} we present the throughput of training the diffusion models at different cluster scales and batch sizes.

For single backbone models (Fig.~\ref{fig:throughput_sd_v21} and~\ref{fig:throughput_cldm_v10}),~\name{} outperforms other pipeline systems both with and without self-conditioning, as it efficiently fills pipeline bubbles with non-trainable layer execution.
When training on a machine, device utilization determines performance.
~\name{} can outperform data parallel baselines because both trainable model stages and the non-trainable part occupy only part of the cluster, and it processes the input batch with a larger local batch size, thus achieving better device utilization.
At batch size 256, ~\name{} achieves 1.44x and 1.16x speedups over GPipe and DeepSpeed, respectively, when training Stable Diffusion v2.1.

When training on multiple machines, synchronization overhead has a more significant impact on training throughput.
\name{} outperforms data parallel baselines as it can mitigate the overhead in two ways:
(1) Each device hosts fewer parameters in pipeline training, so less synchronization communication is required;
(2) Synchronization can be overlapped with the non-trainable part, further reducing its impact on the throughput.
At batch size 2048 on 64 GPUs, ~\name{} achieves 1.41x and 1.28x speedups over GPipe and DeepSpeed training of ControlNet v1.0.

For the cascaded diffusion models (Fig.~\ref{fig:throughput_cdm_lsun} and Fig.~\ref{fig:throughput_cdm_imagenet}),~\name{}'s throughput is comparable to DeepSpeed-P for two reasons:
(1) In both CDM models, there is little non-trainable part to fill bubbles, so we cannot get speedup from the non-trainable part;
(2) Backbone sizes in both CDMs are relatively close to each other, and DeepSpeed-S already achieves balanced training iteration time with respect to backbones.
However, ~\name{} can still achieve a higher training batch size compared to DeepSpeed-P because the activation memory of micro-batches does not persist during the entire backward process.

\subsection{Pipeline bubble ratio}
\begin{figure}[ht]
    \centering
    \vspace{-1.5em}
    \includegraphics[width=0.99\linewidth]{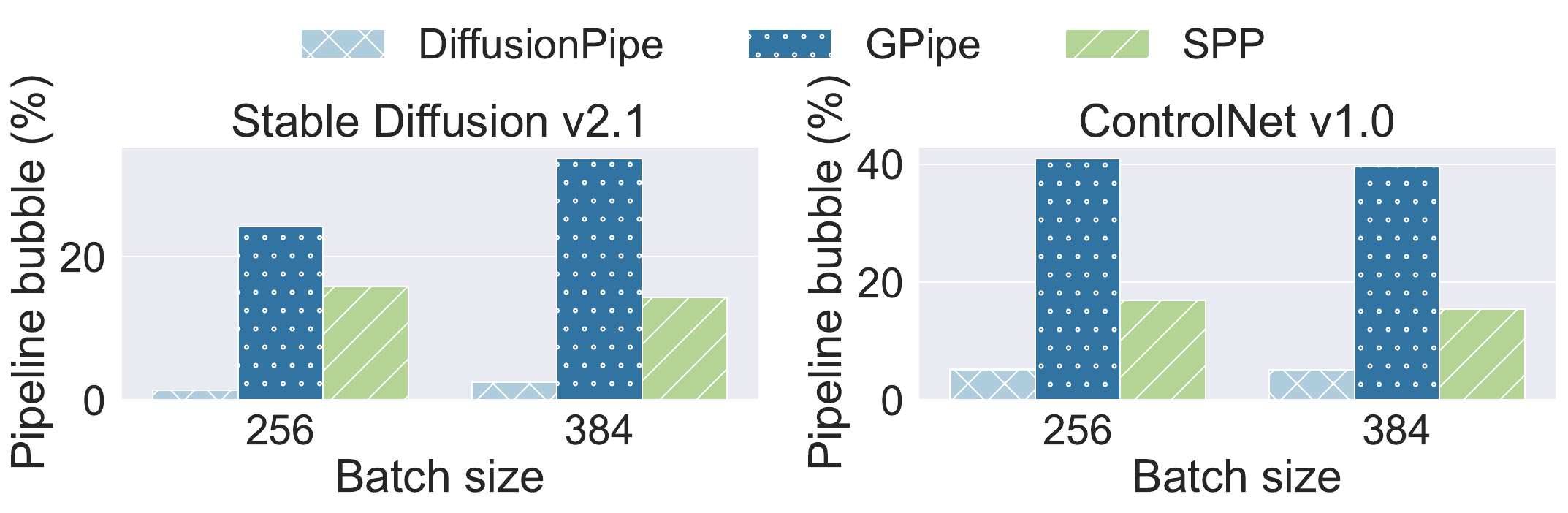}
    \vspace{-1em}
    \caption{Pipeline bubble ratio on 8 GPUs}
    \label{fig:evaluate_pipeline_bubble_ratio}
    \vspace{-1.5em}
\end{figure}

In Fig.~\ref{fig:evaluate_pipeline_bubble_ratio}, we observe that~\name{} can reduce the pipeline bubble ratio to less than 5\% for both Stable Diffusion v2.1 and ControlNet v1.0, which is dramatically lower than other pipeline training baselines.
The unfilled pipeline bubble time can be explained by:
(1) The difference between the actual execution time and the profiled execution time (used to drive the bubble-filling algorithm);
(2) The non-continuous execution times of non-trainable layers, which make it unlikely to perfectly fill the bubble.

\subsection{Ablation study}
\begin{figure}[!h]
    \centering
    \vspace{-1.5em}
    \includegraphics[width=0.99\linewidth]{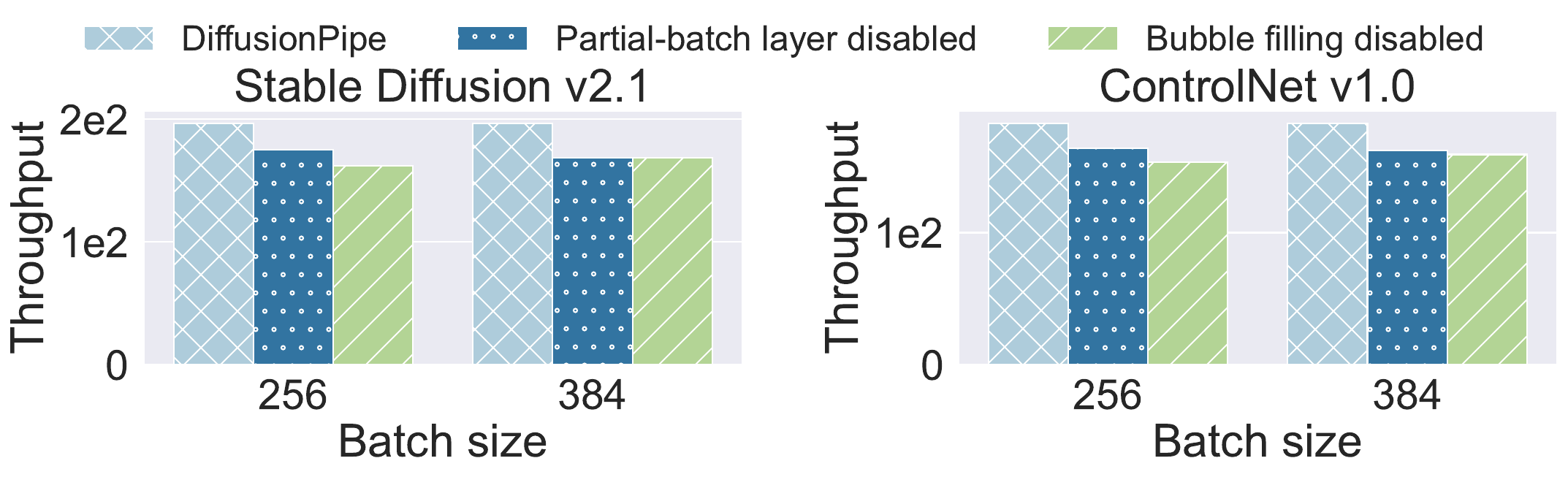}
    \vspace{-1em}
    \caption{Ablation study on 8 GPUs\ (samples/second)}
    \label{fig:ablation}
    \vspace{-1em}
\end{figure}

In Fig.~\ref{fig:ablation}, we evaluate the throughput of~\name{} when the partial-batch layer design is disabled and when the pipeline bubble filling design is completely disabled, respectively.
We observe that disabling the partial-batch layer significantly degrades throughput, and disabling bubble filling degrades it even more (by 10.9\% and 17.6\% for ControlNet v1.0 at batch size 256).
This demonstrates that pipeline bubble filling and the partial-batch layer design can effectively improve training efficiency.
We also observe that at batch size 384, disabling the partial-batch layer achieves almost the same throughput as no bubble filling, indicating that the extra-long layer in Fig.~\ref{fig:execution_times} blocks almost all layers during bubble filling, and validating our partial-batch design.

\subsection{Pre-processing overhead}\label{sec:pre_processing_overhead}
Pre-processing, including profiling, running the model partitioning and the pipeline bubble filling algorithm, is performed once and completes within a few minutes, which is acceptable given training usually takes much longer time.

Profiling is executed in parallel on all GPUs, and its overhead is decided by the number of GPUs.
A typical profiling time of Stable Diffusion v2.1 on 2 AWS EC2 p4de.24xlarge machines at batch size 512 is 55 seconds.

Model partitioning algorithm is executed in parallel on all CPUs in the host machine, and its overhead is decided by the number of CPUs in the host, the number of trainable components and the number of layers in them.
For Stable Diffusion v2.1 and ControlNet v1.0 at the same setting, the overhead is about 0.5 second.

Pipeline bubble fulling algorithm is executed on only 1 CPU, and its overhead is decided by the number of pipeline bubbles and the number of non-trainable components.
For the same models at the same setting, the overhead is less than 1 second.

%% file: sources/conclusion.tex
\section{Conclusion}
This paper presents~\name{}, a system that automatically optimizes pipeline training for large diffusion models.
Our unified partitioning algorithm for the trainable part optimizes partitioning schemes of multiple training scenarios of diffusion models.
We also propose to fill pipeline bubbles with the non-trainable part of diffusion models, which achieves higher training throughput compared to pipelining only the backbone model and training in data parallel.
Experimental results demonstrate that~\name{} achieves speedups of up to 1.41x compared to pipeline baselines and 1.28x compared to data parallel baselines. 
This is accomplished by reducing the pipeline bubble to less than 5\% of the training iteration time. 
Moreover,~\name{} enables the use of larger training batch sizes in comparison to data parallel baselines.
Our design of filling pipeline bubbles with non-trainable parts can extend to more applications, e.g., training or fine-tuning diffusion models with transformer backbones~\cite{Peebles2022ScalableDM,Ma2024SiTEF,Chen2023PixArtFT,Chen2024PIXARTFA}, together with multimodal models with frozen encoder components~\cite{Li2023FinetuningML,Li2023FrozenLM,Yu2023SPAESP}.

%% file: sources/acknowledgement.tex
\section{acknowledgement}
We would like to thank the Program Chairs and anonymous reviewers for their valuable feedback.
This work was supported by an Amazon Research Award (ARA) on AWS AI and grants from Hong Kong RGC under the contracts HKU 17208920, 17204423 and C7004-22G (CRF).